\documentclass[aps,prd,twocolumn,showpacs,groupedaddress,floatfix]{revtex4-1}

\usepackage{graphicx}
\usepackage{dcolumn}
\usepackage{bm}
\usepackage{amsmath}
\usepackage{amssymb}
\usepackage{url}
\usepackage[normalem]{ulem}

\newcommand{\beq}{\begin{eqnarray}}
\newcommand{\eeq}{\end{eqnarray}}

\newcommand{\real}{{\sf I}\kern-.12em{\sf R}}
\newcommand{\comp}{{\sf I}\kern-.50em{\sf C}}
\newcommand{\unity}{{\sf I}\kern-.54em{\sf 1}}

\def\spose#1{\hbox to 0pt{#1\hss}}
\def\ltapprox{\mathrel{\spose{\lower 3pt\hbox{$\mathchar"218$}}
 \raise 2.0pt\hbox{$\mathchar"13C$}}}

\begin{document}

\title{On the critical line of 2+1 flavor QCD}
\author{Paolo Cea}
\affiliation{Dipartimento di Fisica dell'Universit\`a di Bari 
and INFN - Sezione di Bari, I-70126 Bari, Italy}
\email{paolo.cea@ba.infn.it}
\author{Leonardo Cosmai}
\affiliation{INFN - Sezione di Bari, I-70126 Bari, Italy}
\email{leonardo.cosmai@ba.infn.it}
\author{Alessandro Papa}
\affiliation{Dipartimento di Fisica dell'Universit\`a della Calabria \\
and INFN - Gruppo collegato di Cosenza, I-87036 Arcavacata di Rende, 
Cosenza, Italy}
\email{papa@cs.infn.it}

\date{\today}

\begin{abstract}
We determine the curvature of the (pseudo)critical line of QCD with $n_f$=2+1 staggered 
fermions at nonzero temperature and quark density, by analytic continuation 
from imaginary chemical potentials. Monte Carlo simulations are performed 
adopting the HISQ/tree action discretization, as implemented in the code
by the MILC collaboration, suitably modified to include a nonzero imaginary 
baryon chemical potential.
We work on a line of constant physics, as determined in Ref.~\cite{Bazavov:2011nk}, adjusting the couplings so as to keep 
the strange quark mass $m_s$ fixed at  its physical value, with a light 
to strange mass ratio $m_l/m_s=1/20$. 
In the present investigation we set the chemical potential at the
same value for the three quark species, $\mu_l=\mu_s\equiv \mu$.
We explore lattices of different spatial extensions, $16^3\times 6$ and 
$24^3\times 6$, to check for finite size effects, and present results on a 
$32^3 \times 8$ lattice, to check for finite cut-off effects.
We discuss our results for the curvature $\kappa$  of the critical line at $\mu = 0$,
which indicate $\kappa=0.018(4)$,
and compare them with previous lattice determinations by alternative methods 
and with experimental determinations of the freeze-out curve. 
\end{abstract}

\pacs{11.15.Ha, 12.38.Gc, 12.38.Aw}

\maketitle

\section{Introduction}
\label{introd}

It is now well established that Quantum Chromodynamics 
(QCD) is the theory underlying strong interactions. 
As such, it must be able to account for the different phases 
of strongly interacting  matter  under usual or unusual (extreme) conditions.
In particular, a transition or rapid crossover is thought to exist 
from a   low-temperature  hadronic phase to a   high-temperature  
Quark-Gluon Plasma (QGP) phase; 
the line separating these two phases in the
temperature - baryon density plane is called the QCD (pseudo)critical line and
has been the subject of many theoretical investigations.

Determining the exact location of this line and the nature of 
the transition across it has many important theoretical and
phenomenological implications, which
go from the physics of the early Universe, corresponding to the 
high $T$ - low baryon density region of the phase diagram, to the physics
of the interior of some compact astrophysical objects, corresponding
to the low $T$ - high density region.
Moreover, various experiments have been devised or have been planned
in order to study this transition under controlled conditions in a laboratory, 
via heavy-ion collisions at ultrarelativistic energies. 

Depending on the beam energy, different conditions of temperature and baryon 
density can be realized in the fireball produced 
after the collision, such that the QGP phase appears as
a transient state, before the system freezes out and partons recombine into
hadrons. For a given collision energy, the particle yields are found to
be well described by a thermal-statistical model assuming approximate 
chemical equilibrium, as realized at the chemical freeze-out
point, in terms of only two parameters, the freeze-out 
temperature $T$ and the baryon chemical potential $\mu_B$. The set of
freeze-out parameters determined in experiments with different collision
energies lie on a curve in the $(T,\mu_B)$-plane, extending up to
$\mu_B\lesssim$ 800 MeV (see Fig.~1 of Ref.~\cite{Cleymans:2005xv},
or Ref.~\cite{Becattini:2012xb} for a recent re-analysis of experimental data).

There is no compelling reason for the chemical
freeze-out curve and the QCD (pseudo)critical line to 
coincide. Chemical freeze-out is reached, as the fireball
cools down, subsequently to re-hadronization. Hence, the only
assumption that can be made  {\it a priori}  is that  
the freeze-out curve lies below 
the (pseudo)critical line in the $\mu_B$-$T$ plane.
However, a reasonable guess is that chemical freeze-out 
is reached shortly after hadronization, so that the two
curves lie close to each other. In general, the QCD critical line, as well
as the freeze-out curve, can be parameterized, at low baryon densities, by
a lowest order Taylor expansion in the baryon chemical potential, as follows
\begin{equation}
\frac{T(\mu_B)}{T_c}=1-\kappa \left(\frac{\mu_B}{T(\mu_B)}\right)^2\;,
\label{curv}
\end{equation}
where $T_c$ is the pseudo-critical temperature at vanishing baryon density.

Within QCD, a first-principle approach aimed at locating the critical line by
means of numerical simulations on a space-time lattice is unfeasible at 
nonzero baryon density, due to the well known ``sign problem'': the QCD 
fermion determinant becomes complex and the probability interpretation of the 
measure
of the Euclidean path integral, which is necessary for the application of 
standard Monte Carlo importance sampling, is lost.

Several methods have been invented to attack this problem at an algorithmic
level or to circumvent it (for a review, 
see~\cite{Philipsen:2005mj,*Schmidt:2006us,*deForcrand:2010ys,*Aarts:2013bla}): 
reweighting from the ensemble at
$\mu_B=0$~\cite{Barbour:1997bh,*Fodor:2001au},
the Taylor expansion
method~\cite{Gottlieb:1988cq,*Choe:2002mt,*Choe:2001cq,
*Allton:2005gk,*Ejiri:2005uv}, the canonical
approach~\cite{Alford:1998sd,Hasenfratz:1991ax,deForcrand:2006ec},
the density of states
method~\cite{Bhanot:1986kv,*Karliner:1987cu,*Azcoiti:1990ng,*Ambjorn:2002pz}
and the method of analytic continuation from an imaginary chemical
potential~\cite{deForcrand:2002ci,*deForcrand:2003hx,Lombardo:1999cz,*D'Elia:2002gd,*D'Elia:2004at,Azcoiti:2005tv,deForcrand:2006pv,*deForcrand:2008vr,*deForcrand:2007rq,Cea:2010md,Cea:2010fh,*Cea:2012vi,Cea:2012ev,Wu:2006su,Nagata:2011yf,Giudice:2004se,*Giudice:2004pe,Papa:2006jv,Cea:2007wa,*Cea:2007wa,*Cea:2010bp,*Cea:2009ba,*Cea:2010bz,Karbstein:2006er}.

A comparison among different approaches has been possible only in few 
cases. QCD with $n_f=4$ in the standard staggered formulation,
discretized on $N_t = 4$ lattices with a bare quark mass $am = 0.05$,
has been the laboratory for many investigations: in that
case the transition is first order all the way along the critical line, 
and all methods agree in the range $\mu_B/(3 T)\lesssim1$ 
(see Refs.~\cite{deForcrand:2006ec,Fodor:2012rma} and 
Fig.~8 of Ref.~\cite{Cea:2010md}). No such direct comparisons exist
for $n_f = 2$ QCD, however different discretizations with unphysical
quark masses lead to compatible results
(see, {\it e.g.},  the discussion in Sect.~3 of Ref.~\cite{Cea:2012ev}).

The situation in QCD with $n_f=2+1$ and physical or almost physical
quark masses deserves a
more detailed discussion. Here it is widely accepted now that the transition 
at $\mu_B=0$ is a smooth crossover~\cite{Aoki:2009zzc}, thus implying that the 
determination of the transition temperature $T_c$ and the curvature of the 
critical line depend on the observable adopted to probe the transition. 
Indeed, with 
particular reference to the curvature $\kappa$ defined in Eq.~(\ref{curv}), the
Budapest-Wuppertal collaboration~\cite{Endrodi:2011gv}, using a Symanzik
improved gauge action and stout-link improved staggered fermions on
lattices with temporal size $N_t=6,8,10$ and aspect ratios equal to 
three and four, finds, after continuum extrapolation, 
$\kappa=0.0089(14)$ by the Taylor
expansion method with the strange quark number susceptibility as probe
observable and $\kappa=0.0066(20)$ when, instead, the renormalized chiral 
condensate is used. The Bielefeld-BNL collaboration~\cite{Kaczmarek:2011zz}, 
using the p4-action on lattices with $N_t=4$ and 8, and aspect ratios
up to four, finds $\kappa=0.0066(7)$
again with the Taylor expansion method and the light quark susceptibility as
a probe observable. Another collaboration~\cite{Falcone:2010az} adopted
improved staggered fermions in the p4fat3 version, on lattices with $N_t=4$
and aspect ratio four with physical strange quark mass and pion mass at 220 
MeV, getting $\kappa=0.0100(2)$ by the method of analytic 
continuation, with the Polyakov loop phase as a probe.

The comparison of these results for the curvature $\kappa$ with those
obtained for the freeze-out curve and 
coming from the experiments with heavy-ion collisions is far from being
satisfactory. According to the analysis of Ref.~\cite{Cleymans:2005xv},
the curvature $\kappa$ of the freeze-out curve is a factor two to three higher
than the above lattice determinations, even if a recent 
reanalysis~\cite{Becattini:2012xb}, which includes the effects
of inelastic collisions taking place after freeze-out, seems to 
reduce the value of $\kappa$, bringing it in a better agreement with 
existing lattice results.

In such a situation a new, independent lattice determination of the QCD 
critical line at small baryon densities could provide us with useful additional 
information and help us identifying possible sources of systematic 
uncertainties in the theoretical determination of the (pseudo)critical line.
Indeed, while systematic effects within each single method trying 
to circumvent the sign problem may seem to be well under control,
it is only the comparison between different independent methods which
could provide a clear, final picture.

The aim of this work is to have a first estimate of the QCD critical line by the method of 
analytic continuation, using the HISQ/tree action of the MILC collaboration 
with 2+1 staggered fermions, properly modified to be endowed with an
imaginary chemical potential, common to each fermion. The strange mass is set 
at the physical value and simulations are performed on the line of constant 
physics (LCP) with the light quark mass fixed at $m_l=m_s/20$, as determined 
in Ref.~\cite{Bazavov:2011nk}.  As for quark chemical potentials, in the
present study we assign the same value to the three quark species, 
$\mu_l=\mu_s\equiv \mu$. This choice is certainly the most convenient for 
studying the theory at imaginary chemical potentials, since it leads to
a simpler structure of the phase diagram for negative values of $\mu^2$.
For the comparison with the freeze-out curve, also settings
with $\mu_s \neq \mu_l$ should be taken into account. We plan to
consider this issue in forthcoming studies.
We explore lattices of different
spatial extensions, $16^3\times 6$ and $24^3\times 6$, 
to check for finite size effects, and present 
results on a $32^3 \times 8$ lattice,
to check for finite cut-off effects.

\section{Simulation details and numerical results}

We perform simulations of lattice QCD with 2+1 flavors of rooted staggered 
quarks at imaginary quark chemical potential.
We have made use of the HISQ/tree action~\cite{Follana:2006rc,Bazavov:2010ru} 
as implemented in the publicly available MILC code~\cite{MILC}, 
which has been suitably modified by us in order to introduce an imaginary quark
chemical potential $\mu = \mu_B/3$. 
That has been done by multiplying all forward and backward 
temporal links entering the discretized Dirac operator 
by $\exp (i a \mu )$ and $\exp (- i a \mu )$, respectively: in this way,
the fermion determinant is still real and positive, so that standard
Monte Carlo methods can be applied.
As already remarked above, in the present study we have 
$\mu=\mu_l=\mu_s$.
All simulations make use of the rational hybrid Monte Carlo (RHMC) 
algorithm. The length of each RHMC trajectory has been set to  
$1.0$ in molecular dynamics time units.

\begin{figure}[tb]
\includegraphics*[width=0.8\columnwidth]
{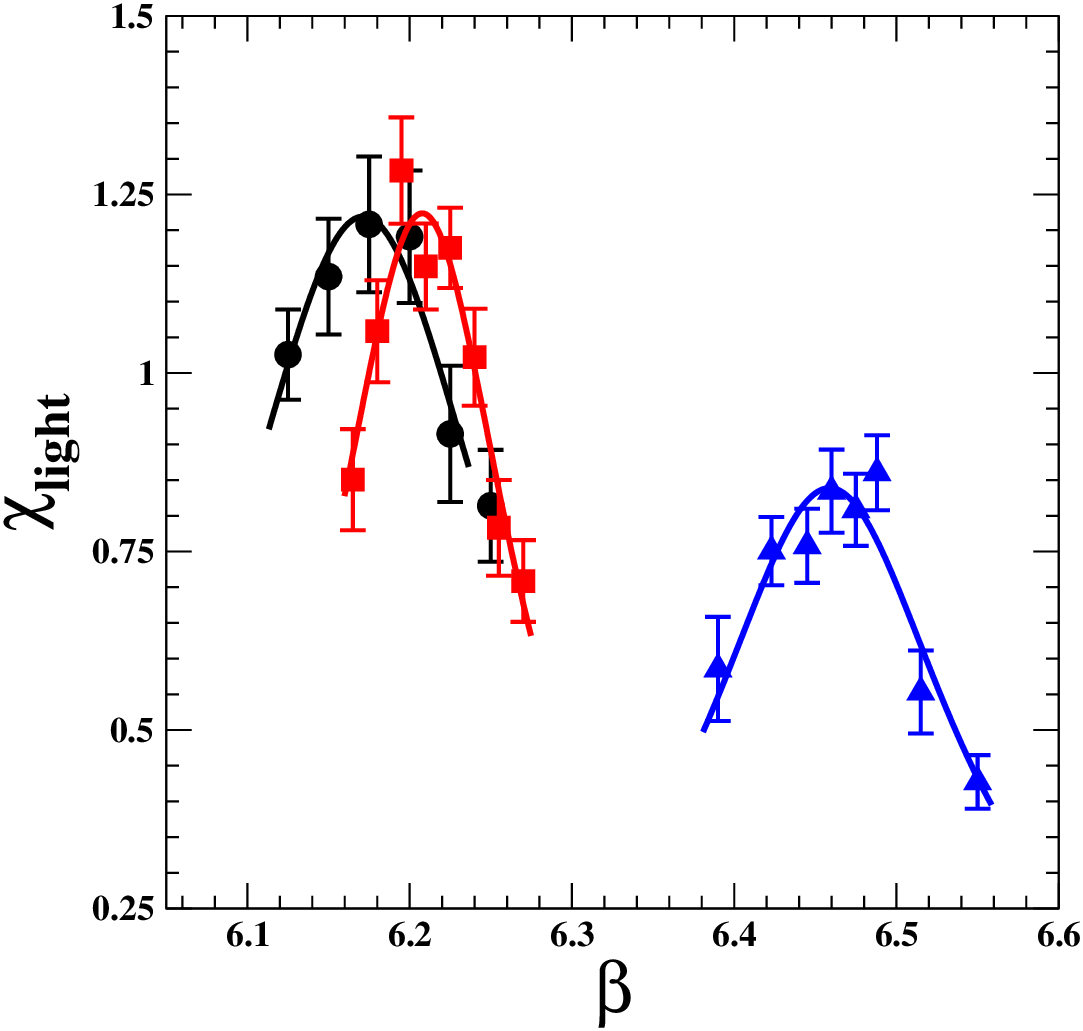}
\caption{The real part of the disconnected susceptibility of the light quark 
chiral condensate for $16^3\times 6$ and $24^3\times 6$
(full circles and full squares, respectively) and for $32^3 \times 8$ 
(full triangles) at 
$\mu/(\pi T)=0.2i$. Full lines are the fits to the peaks using a Lorentzian.}
\label{fig_chiral_light_suscep}
\end{figure}

\begin{table}[tb]
\setlength{\tabcolsep}{1pc}
\centering
\caption[]{Summary of the values of the critical couplings $\beta_c$ for 
the imaginary quark chemical potentials $\mu$ considered in this work.
The data for $\mu=0$ on $24^3\times6$ lattice and on $32^3\times6$ lattice
have been estimated from the disconnected chiral susceptibilities reported respectively on
Table X and Table XI of Ref.~\cite{Bazavov:2011nk}.}
\begin{tabular}{clll}
\hline
\hline
lattice & $\mu/(\pi T)$ & $\beta_c$ & $T_c(\mu)/T_c(0)$  \\
\hline
16$^3\times 6$ & 0.      & 6.102(8)  & 1.000 \\
               & 0.15$i$ & 6.147(10) & 1.045(13) \\
               & 0.2$i$  & 6.171(12) & 1.070(15) \\
               & 0.25$i$ & 6.193(14) & 1.093(17) \\
\hline
24$^3\times 6$ & 0.      & 6.148(8)  & 1.000 \\
               & 0.2$i$  & 6.208(5)  & 1.060(10) \\
\hline
32$^3\times 8$ & 0.      & 6.392(5)  & 1.000 \\
               & 0.2$i$  & 6.459(9)  & 1.068(11) \\
\hline
\hline
\end{tabular}
\label{summary}
\end{table}

We have simulated QCD at finite temperature and imaginary quark chemical potential 
on lattices of size $16^3\times 6$, {$24^3 \times 6$ and $32^3 \times 8$.}
{In particular, most simulations have been performed on the smallest lattice, while
for} $\mu/(\pi T)=0.2i$ we have 
considered also a $24^3 \times 6$ lattice and a  $32^3 \times 8$ lattice,
{in order to check for finite size and for finite cut-off effects.}
We have discarded typically not less than one thousand trajectories for each 
run and have collected from {4k to 8k} trajectories for measurements.

The pseudocritical  line $\beta_c(\mu^2)$  has been determined as the value for which 
the disconnected susceptibility of the light quark
chiral condensate exhibits a peak. To precisely localize the peak, a 
Lorentzian fit has been used. 
For the $24^3\times6$ and  $32^3\times8$ lattices,
the values of the susceptibility at $\mu/(\pi T)=0$ have been
taken from Table~X and Table~XI of Ref.~\cite{Bazavov:2011nk}, respectively
(see Table I for the fitted pseudo-critical couplings). 
For illustrative purposes, 
in Fig.~\ref{fig_chiral_light_suscep} we display our determination of the 
pseudo-critical couplings at $\mu/(\pi T)=0.2i$ for $16^3\times6$, $24^3\times6$, and $32^3\times8$ lattices. 
We notice that the discrepancy in the determination of $\beta_c$ on 
the $16^3 \times 6$ and the $24^3 \times 6$ lattices, which may indicate the presence
of finite size effects, will be strongly suppressed when considering the ratio of temperatures,
$T_c(\mu)/T_c(0)$.

To determine the ratio $T_c(\mu)/T_c(0)$ we need to set the lattice spacing.
This is done following the discussion in Appendix B of 
Ref.~\cite{Bazavov:2011nk}, where, for this particular value of $m_l/m_s$, 
the spacing is given in terms of the $r_1$ parameter:
\begin{equation}
\label{scale}
\frac{a}{r_1}(\beta)_{m_l=0.05m_s}=
\frac{c_0 f(\beta)+c_2 (10/\beta) f^3(\beta)}{
1+d_2 (10/\beta) f^2(\beta)} \; ,
\end{equation}
with $c_0=44.06$, $c_2=272102$, $d_2=4281$, $r_1=0.3106(20)\ {\text{fm}}$~\cite{Bazavov:2010hj}
and
\begin{equation}
\label{beta function}
f(\beta)=(b_0 (10/\beta))^{-b_1/(2 b_0^2)} \exp(-\beta/(20 b_0))\;,
\end{equation}
where $b_0$ and $b_1$ are the  universal coefficients of the two-loop beta 
function. 
\begin{figure}[tb]
\includegraphics*[width=0.8\columnwidth]
{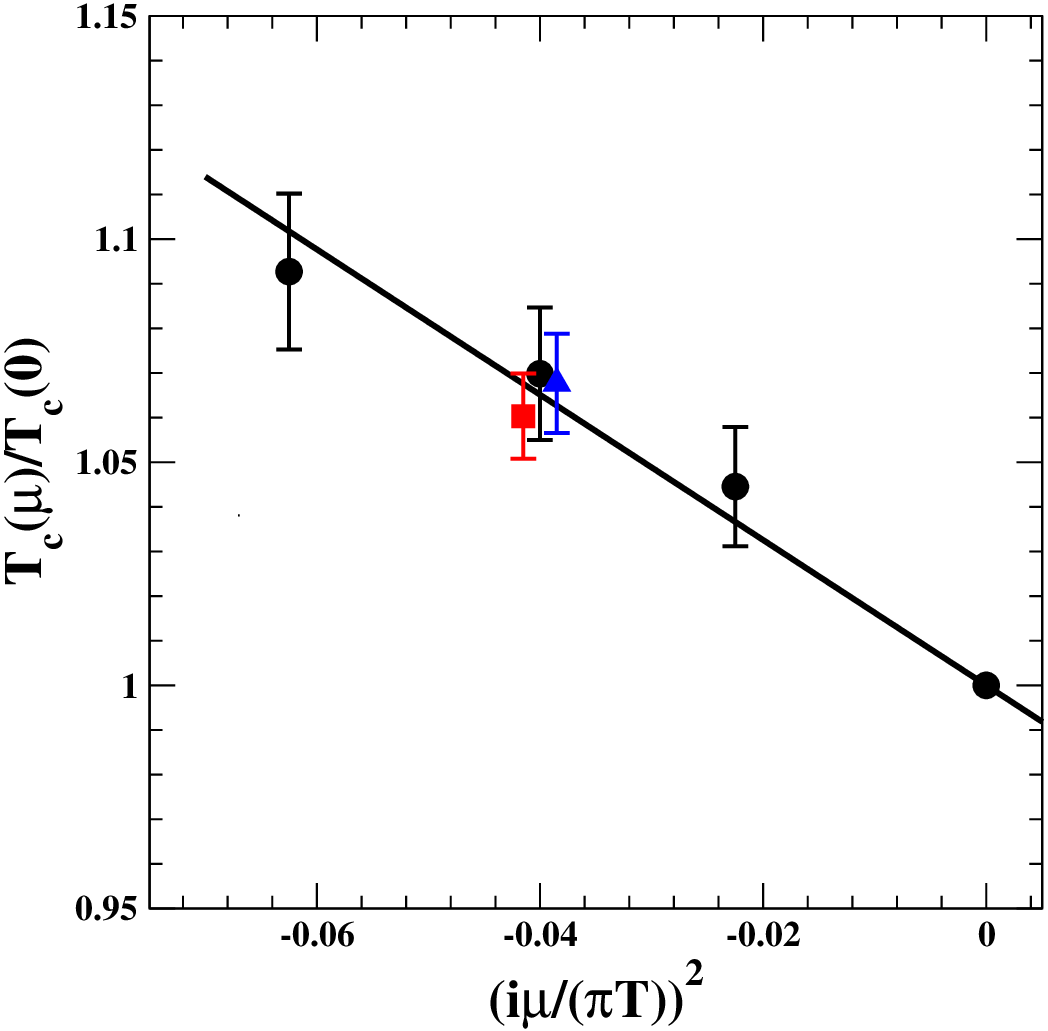}
\caption{$T_c(\mu)/T_c(0)$ versus $((i\mu)/(\pi T))^2$ obtained on a $16^3\times6$ lattice (full circles), on a $24^3\times6$  lattice (full square) and on a $32^3\times8$ lattice (full triangle).
For the sake of readability the abscissae at $((i\mu)/(\pi T))^2=-0.04$  for $24^3\times6$ and $32^3\times8$ data have been slightly shifted.
The full line is a linear fit to the data on the $16^3\times6$ lattice.
}
\label{fig_slope}
\end{figure}

\begin{figure}[t!]
\includegraphics*[width=0.99\columnwidth]
{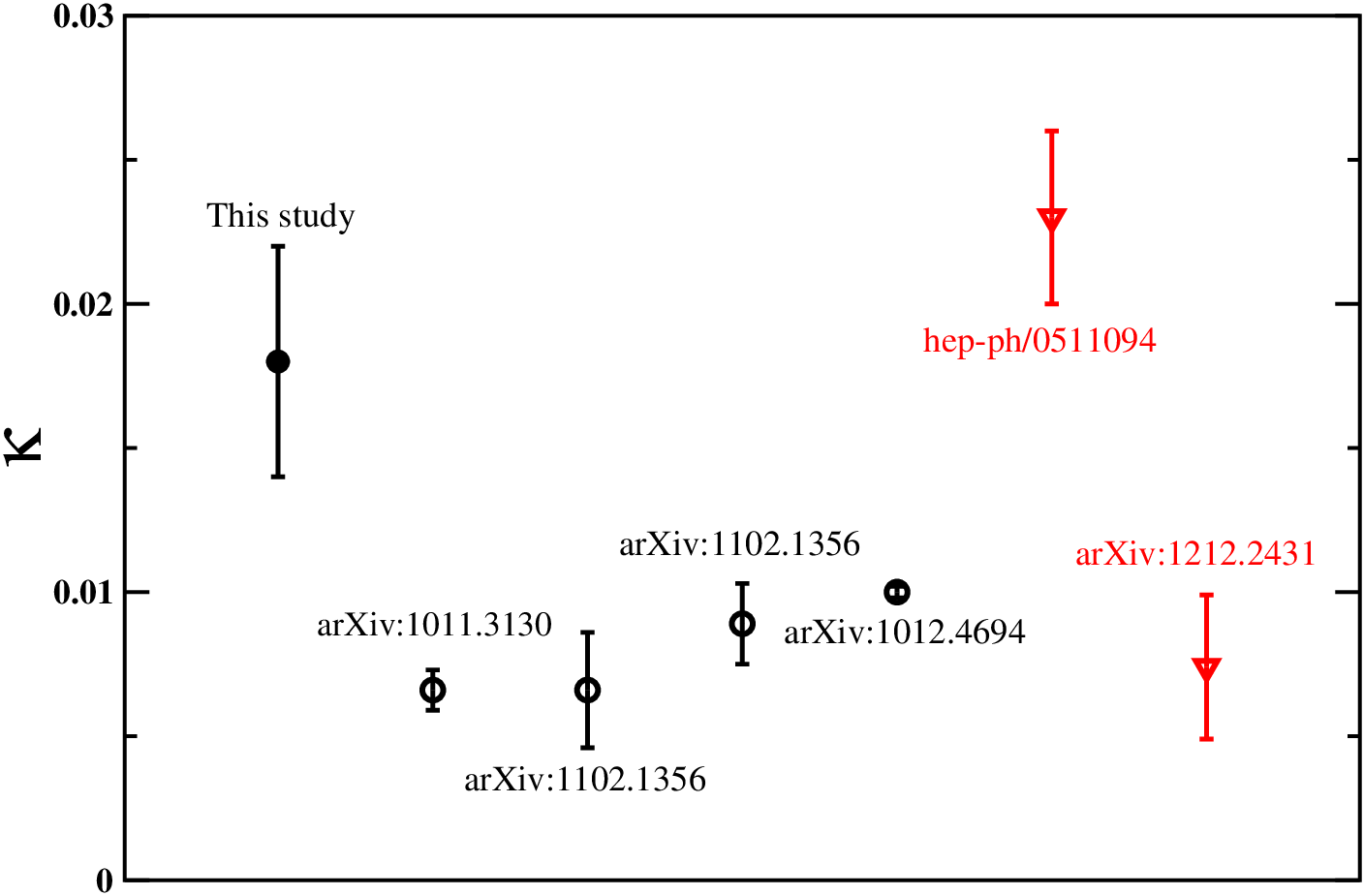}
\caption{Comparison of different determinations of 
the curvature of the chemical freeze-out curve and of the 
pseudo-critical line for QCD with $n_f = 2+1$.
From left to right: {\em i)} analytic continuation, disconnected chiral 
susceptibility, this study; 
{\em ii)} Taylor expansion, chiral susceptibility, Ref.~\cite{Kaczmarek:2011zz}; 
{\em iii)} Taylor expansion, chiral condensate, Ref.~\cite{Endrodi:2011gv}; 
{\em iv)}  Taylor expansion, strange quark number susceptibility, Ref.~\cite{Endrodi:2011gv}; 
{\em v)} analytic continuation, Polyakov loop, Ref.~\cite{Falcone:2010az}; {\em vi)} freeze-out curvature,
standard analysis, Ref.~\cite{Cleymans:2005xv}; 
{\em vii)} freeze-out curvature, revised analysis of Ref.~\cite{Becattini:2012xb}.
}
\label{k_compare}
\end{figure}

From $a(\beta)$ we determine, for each explored lattice size separately, 
$T_c(\mu)/T_c(0) = a(\beta_c(0))/a(\beta_c(\mu))$. Data for 
$T_c(\mu)/T_c(0)$ versus $\mu/(\pi T)$ are reported in Fig.~\ref{fig_slope}. 
For the $16^3\times6$ lattice, where the determination at three different
values of $\mu$ is available, we have a tried a linear fit  in $\mu^2$:
\begin{equation}
\label{linearfit}
\frac{T_c(\mu)}{T_c(0)} = 1 + R_q \left(\frac{i \mu}{\pi T_c(\mu)}\right)^2 \;,
\end{equation}
which works well over the whole explored range ($\chi^2/{\rm d.o.f.} = 0.39$) 
and gives us access to the curvature $R_q$. On the other lattices, assuming 
that linearity in $\mu^2$ still holds, we can extract $R_q$ from the 
determination at $\mu/(\pi T)=0.2i$; we notice that such an assumption, for 
the given value of $\mu$, is consistent with all previous studies on the 
systematics of analytic 
continuation~\cite{Cea:2010md,Cea:2010fh,*Cea:2012vi,Cea:2012ev}. 
Our results are:
\begin{eqnarray}
\label{Rq}
R_q(16^3\times6) & =& -1.63(22) \,, \quad \kappa =  0.0183(24) \,, \nonumber \\
R_q(24^3\times6) & =& -1.51(25) \,, \quad \kappa =  0.0170(28) \,,  \\
R_q(32^3\times8) & =& -1.70(29) \,, \quad \kappa =  0.0190(32) \,, \nonumber
\end{eqnarray}
where $\kappa = -R_q/(9 \pi^2)$  is the curvature parameter introduced 
in Eq.~(\ref{curv}).
The results provide evidence that finite size and finite cut-off
systematic effects are within our present statistical uncertainties.
We cannot yet try an extrapolation to the continuum limit of our results,
however, taking into account the statistical errors and the observed variations
of the results with the lattice size and the ultraviolet cutoff, our present
estimate for kappa is
\begin{equation}
\label{curvature}
\kappa = 0.018(4) \;.
\end{equation}

\section{Conclusions and discussion}

We have presented the first results of our study of QCD with $n_f = 2+1$ 
flavors discretized in the HISQ/tree rooted staggered fermion formulation
and in the presence of an imaginary baryon chemical potential, with a physical
strange quark mass and a light to strange mass ratio $m_l/m_s = 1/20$, 
and $\mu=\mu_l=\mu_s$.

Our main
result is an estimate of the curvature of the pseudocritical line
in the temperature - baryon chemical potential, defined in Eq.~(\ref{curv}), 
which has been obtained by the method of analytic continuation. 

\begin{figure}[tb]
\includegraphics*[height=0.295\textheight,width=0.95\columnwidth]
{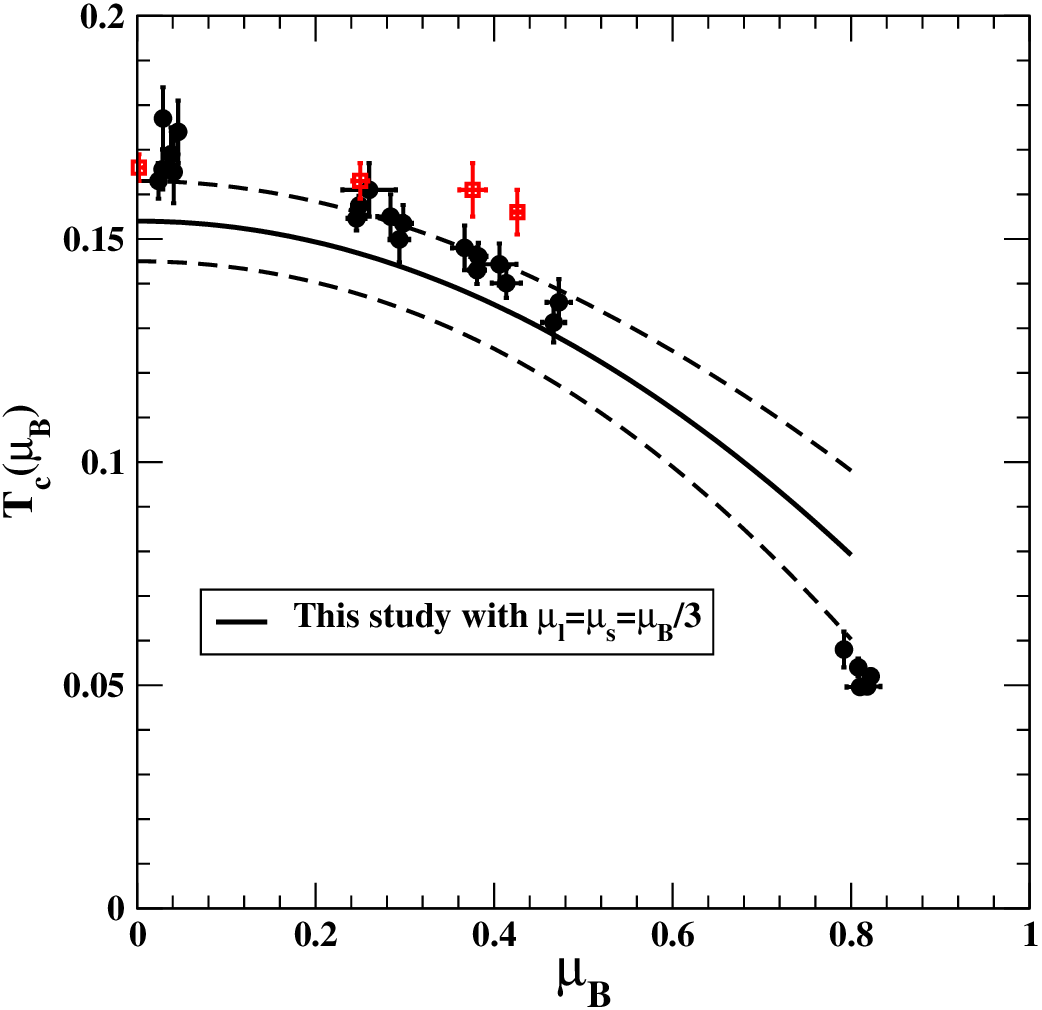}
\caption{$T_c(\mu_B)$ versus $\mu_B$ (units in GeV). Experimental values of $T_c(\mu_B)$ are taken
from Fig.~1 of Ref.~\cite{Cleymans:2005xv} (circles) and from 
Table I of Ref.~\cite{Becattini:2012xb} (squares) 
respectively for the standard and the modified 
statistical hadronization model.
 The solid line is a parametrization corresponding to
$T_c(\mu_B) = T_c(0) - b \mu_B^2$ with $T_c(0)=0.154(9)\, \text{GeV}$ and $b=0.117(27)\,{\text{GeV}}^{-1}$. The dashed lines represent  the corresponding 
error band.}
\label{Tcmu}
\end{figure}

It is interesting to compare our estimate with previous lattice results, which 
have been mostly obtained by the Taylor expansion method, and with the 
estimates of the freeze-out curve. Such a comparison is performed in 
Fig.~\ref{k_compare}.
We stress once again that our investigation has been performed with 
$\mu_l=\mu_s$, as in the numerical setup of 
Ref.~\cite{Falcone:2010az}, while the other results in 
Fig.~\ref{k_compare} have been obtained for $\mu_s=0$.

Regarding the freeze-out curve, we report two different estimates. The first
is from the analysis of Ref.~\cite{Cleymans:2005xv}, which is based
on the standard statistical hadronization model; there the authors
parametrize the freeze-out curve as 
\begin{equation}
\label{cleymans}
T_c(\mu_B) = a - b \mu_B^2 - c \mu_B^4\;,
\end{equation}
with $a=0.166(2) \ {\text{GeV}}$, $b=0.139(16) \ {\text{GeV}}^{-1}$, 
and $c=0.053(21) \ {\text{GeV}}^{-3}$, from which we have derived the 
$\kappa$ value reported in the figure. 
The second estimate is based on 
the estimates for the freeze-out points which are reported in Table I
of Ref.~\cite{Becattini:2012xb} and are based on a modified statistical
reanalysis of the experimental data which includes the effects
of inelastic collisions taking place after freeze-out.

We also report, in Fig.~\ref{Tcmu}, an estimate of the pseudo-critical line 
which is based on our determination of the curvature. Regarding the value
of $T_c$ at $\mu_B = 0$, which is affected by larger finite 
size and finite cutoff effects than $\kappa$, we refer directly to the 
presently accepted continuum extrapolated value, 
$T_c \sim 155$~\cite{Aoki:2006br,*Aoki:2009sc,*Borsanyi:2010bp,Bazavov:2011nk},
and in particular to the one obtained in Ref.~\cite{Bazavov:2011nk}
with the same action adopted in our study, 
$T_c(0) = 154(9)$ MeV. From that and from 
$\kappa = 0.018(4)$ we obtain 
$b=0.117(27)\,{\text{GeV}}^{-1}$ (see Eq.~\ref{cleymans}). 
Freeze-out determinations from 
Refs.~\cite{Cleymans:2005xv, Becattini:2012xb} are reported as well.

Our result for the curvature is typically between  two and three
standard deviations
larger than previous lattice determinations and seems in a better agreement
with the freeze-out curvature based on the standard statistical hadronization
model.

Possible reasons for the disagreement with previous lattice
determinations can lie in the different methods adopted
to avoid the sign problem, in the different lattice
discretizations, as well as in the different observables
used to locate the transition point, and in the setup of 
quark chemical potentials. In this respect, it would very
interesting in the future to redetermine the curvature $\kappa$
using different combinations of methods and lattice discretizations,
such as implementing the Taylor expansion method with the
HISQ/tree discretization, or the method of imaginary chemical
potential with lattice discretizations adopting stout smearing improvement.

Let us conclude by discussing the possible sources of systematic effects
in our estimate. One of them is related to the extrapolation from
imaginary to real chemical potentials: in the case of the $16^3\times 6$
lattice we have performed simulations at different values of imaginary
$\mu$, thus verifying that a linear interpolation (in $\mu^2$)
of data works well. For the other two lattices, instead, we have considered
only one value of imaginary $\mu$ ($\mu/(\pi T)=0.2i$) and the linear
behavior has been assumed. Our previous studies based on analytic
continuation, however, indicate that the chosen value of $\mu$ should
lie well inside the region of linearity. Nevertheless,
we plan to perform a more systematic study of this issue.
Finally, we have verified that finite size and cutoff effects are under
control, within the present statistical accuracy. Still, the
extrapolation to the continuum limit, as well as the extension to the physical
value of the light to strange mass ratio, $m_l/m_s \sim 1/28$, and the 
possible effect of varying the strange quark chemical potential, deserve
further investigations and will be the subject of forthcoming works.

\section*{Acknowledgments}
We thank Massimo D'Elia for collaboration in early stage of this work. 
This work was in part based on the MILC collaborationÕs public lattice gauge theory code. See http://physics.utah.edu/~detar/milc.html.
We acknowledge useful suggestions and help  from  MILC collaboration and especially from Carleton DeTar.
This work has been partially supported by the INFN SUMA project.
Simulations have been performed on BlueGene/Q at CINECA (CINECA-INFN agreement
under INFN project PI12), on the BC$^2$S cluster in Bari and on the CSNIV Zefiro cluster in Pisa.

%

\end{document}